# Bridging the Gap: Leveraging Retrieval-Augmented Generation to Better Understand Public Concerns about Vaccines

Muhammad Javed*

Epidemiology Informatics, Centre for Health Analytics, Murdoch Children's Research Institute (MCRI), Parkville, Victoria, Australia, muhammad.javed@mcri.edu.au

Sedigh Khademi Habibabadi

Epidemiology Informatics, Centre for Health Analytics, Murdoch Children's Research Institute (MCRI), Parkville, Victoria, Australia sedigh.khademi@mcri.edu.au

Christopher Palmer

Epidemiology Informatics, Centre for Health Analytics, Murdoch Children's Research Institute (MCRI), Parkville, Victoria, Australia chris.palmer@mcri.edu.au

Hazel Clothier

Epidemiology Informatics, Centre for Health Analytics, Murdoch Children's Research Institute (MCRI), Parkville, Victoria, Australia jo.hickman@mcri.edu.au

Jim Buttery

Epidemiology Informatics, Centre for Health Analytics, Murdoch Children's Research Institute (MCRI), Parkville, Victoria, Australia.  Infectious Disease Unit, Department of General Medicine, Royal Children's Hospital, Parkville, Australia, Department of Paediatrics, The University of Melbourne, Parkville, Australia, jim.buttery@mcri.edu.au

Gerardo Luis Dimaguila

Health Informatics Group, Murdoch Children's Research Institute (MCRI), Parkville, Australia, gerardoluis.dimaguil@mcri.edu.au

Vaccine hesitancy threatens public health, leading to delayed or rejected vaccines. Social media is a vital source for understanding public concerns, and traditional methods like topic modelling often struggle to capture nuanced opinions. Though trained for query answering, large Language Models (LLMs) often miss current events and community concerns. Additionally, hallucinations in LLMs can compromise public health communication. To address these limitations, we developed a tool (VaxPulse Query Corner) using the Retrieval Augmented Generation technique. It addresses complex queries about public vaccine concerns on various online platforms, aiding public health administrators and stakeholders in understanding public concerns and implementing targeted interventions to boost vaccine confidence. Analysing 35,103 Shingrix® social media posts, it achieved answer faithfulness (0.96) and relevance (0.94).

---

\* Corresponding Author.

# 1 INTRODUCTION

SAEFVIC (Surveillance of Adverse Events Following Vaccination in the Community) monitors vaccine safety in Victoria, Australia. It detects, assesses, and analyses adverse events following immunisation (AEFI). SAEFVIC offers educational resources for the public and healthcare providers, including eLearning and current immunisation guidelines. To engage the community, SAEFVIC publishes weekly vaccine safety reports [1].

Effective communication between healthcare providers and consumers is crucial for health literacy. Social media platforms empower public health organisations to engage diverse audiences and provide accessible, real-time health information [2]. Vaccine hesitancy, a top 10 global health threat, is driven by factors like misinformation and inadequate information on vaccine safety and efficacy [3]. Researchers have employed techniques like topic modelling to analyse public concerns on social media but struggle with complex questions like "Do people believe vaccines cause long-term side effects, and if so, what are these?"

LLMs have revolutionised NLP, including topic modelling and question answering [4]. Pre-trained on large datasets, LLMs are prone to hallucinations, generating inaccurate information that can pose risks in fields like law and medical consultation [5]. To overcome these limitations, Retrieval Augmented Generation (RAG) is a more suitable approach for tasks requiring precise information retrieval and query answering [6].

## 1.1 Objective

In this paper, we are introducing VaxPulse Query Corner (VaxPulse QC), a tool that is based on the RAG technique to answer complex queries about vaccinations. This tool will help us gain a better understanding of public concerns related to vaccinations from online media in near real-time. It is being developed as part of our framework VaxPulse, which is developed to know public sentiments and topics of discussion about vaccinations by processing online and social media data from various platforms in near-real time. VaxPulse QC aims to bridge the gap presented by the limitations of sentiment analysis and topic modelling. This comprehensive solution could also support communication between the public and public health professionals and other stakeholders. By understanding the information gaps and persistent concerns expressed in comments, public health officials can adjust their communication strategies to be more informative and address specific needs.

# 2 Related Work

RAG techniques are leveraged to unlock real-time insights from data streams across diverse fields, enabling faster and more comprehensive analysis. To our knowledge, the RAG method has not been used in vaccination-related online discussions to answer the queries of public health administrators and other stakeholders. Table 1 presents some implementations of the RAG method in the medical domain to highlight its advantages.

**Table 1: RAG implementation in the medical domain**

| Authors | Implementations of the RAG method in the medical domain |
|---|---|
| Kim et al. [7] | Developed the QA-RAG chatbot for the pharmaceutical industry, combining generative AI and RAG to answer user queries using guideline documents. It extracts documents based on LLM-generated answers and queries, reranks them for relevance, and uses ChatGPT-3.5-turbo with few-shot prompting for final responses. |
| Miao et al. [8] | Applied the RAG pipeline in nephrology to address LLM hallucinations by integrating external data for clinical decision-making and educational applications related to kidney disease. |
| Ge et al. [9] | Created LiVersa, a RAG-based system for liver disease queries, using guidelines from the American Association for the Study of Liver Diseases and comparing its responses with trainees' knowledge. |
| Al Ghadban et al. [10] | Employed RAG to enhance healthcare information in low- and middle-income countries, assisting community health workers with maternal care through locally relevant information. They achieved scores of 0.16 in context relevancy, 0.93 in faithfulness, 0.93 in answer relevancy, and 0.26 in context recall. |
| Saba et al. [11] | Used RAG to summarise electronic health records, reducing screen time for patients and medical professionals by extracting and summarising documents with a confidence score. |
| Zakka et al. [12] | Introduced Almanac, an LLM framework with RAG functionality for medical guidelines. Almanac was evaluated against conventional LLMs with 314 clinical questions across nine specialties, showing significant improvements in accuracy and user satisfaction. |

## 3 METHOD

The VaxPulse QC pipeline is illustrated in Figure 1. To clarify our method, we have divided it into four key steps: dataset creation, first iteration, second iteration, and answer formulation.

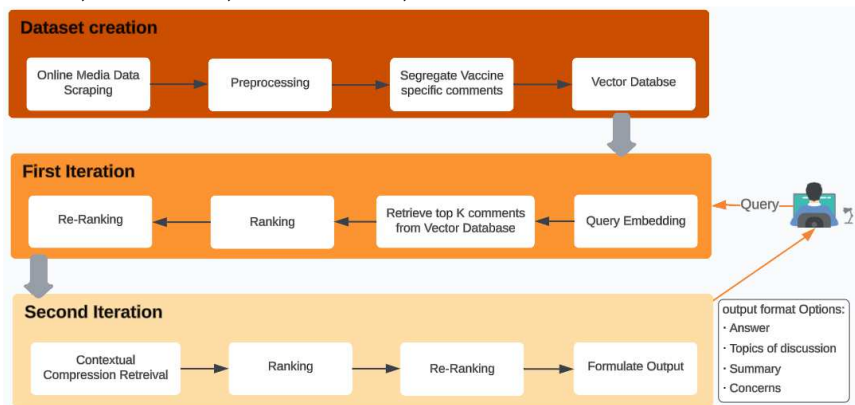

Figure 1: Method

## 3.1 Dataset Creation

We created the dataset using the following steps. Our method paper [13] provides further details on data creation, and we also utilised initial steps in [14] for social bot detection.

### 3.1.1 Online media data scraping

We collect vaccination-related posts from X (former Tweeter), Reddit, YouTube, and Facebook using authorised APIs Tweepy, PRAW, and the Google API Client Library.

### 3.1.2 Preprocessing

Perform certain preprocessing steps to remove unnecessary text from the comments.

### 3.1.3 Segregate Vaccine-specific Comments

We developed a two-step process to identify vaccine-related comments. First, a fine-tuned BERT model categorises comments into three buckets: vaccine (general vaccine information, news, or promotions), personal health mentions, and other (non-vaccine-related discussion). Second, a GPT-4o model with prompt engineering extracts comments about a specific vaccine, such as Shingrix®. This enables public health administrators to address questions like "What is the public sentiment about the second dose of the Shingrix vaccine?"

### 3.1.4 Create embedding and Save in Vector database

Embedding databases significantly enhance RAG models by enabling faster retrieval of relevant information through vector comparisons. This semantic similarity approach ensures focus on relevant content, leading to a more accurate and coherent context. We used OpenAI's text-embedding-ada-002 model to create 1536-dimensional embeddings for our selected dataset, stored in a vector database using the LangChain [15] library.

## 3.2 First Iteration

### 3.2.1 Top K Comments Retrieval

Semantic similarity captured by query embeddings allows the RAG model to retrieve highly relevant documents, improving accuracy. We used OpenAI's text-embedding-ada-002 model for query embedding and LangChain to retrieve the top k (percentage) relevant comments by semantic similarity search. This iteration retrieves a large number of comments from our growing vector database, fed by weekly vaccine-specific data extraction.

### 3.2.2 Ranking & Re-Ranking

LLMs struggle to process information located in the middle of longer contexts. This "lost in the middle" phenomenon affects RAG models, causing them to overlook crucial information within retrieved documents. Re-ordering retrieved documents is a potential solution to mitigate this issue. We employ a two-step ranking process: initial ranking and re-ranking. The initial ranking leverages LangChain's longcontextreorder module to automatically re-order retrieved documents. This places the most similar documents at the top, the least similar in the middle, and the next few at the bottom. To consolidate dispersed information in retrieved documents, we use a multi-stage re-ranking process, applying FlashRank [16] with efficient models like ms-marco-MiniLM-L-12-v2 for smaller sets and rank-T5-flan for larger ones. For fewer than 10 comments, we skip re-ranking, implementing an "early exit" strategy.

## 3.3 Second Iteration

This iteration refines comment selection for accurate results. We prioritise quality over latency by extracting the most relevant information. While the initial iteration extracted documents from a large database, this iteration focuses on a subset of highly relevant comments. To reduce the LLM's cognitive load, we employ contextual compression using LangChain. This technique shortens comments while preserving essential context. By setting an 80% similarity threshold, we select 50% of the top-ranked comments from the retrieved data in first iteration, further improving information quality. We applied ranking and re-ranking steps on the compressed comments in a similar way as discussed in the first iteration.

## 3.4 Formulate Answer

To generate a presentable result for the user by considering the query, ranked chunks of the comments, and a selected option for the output (i.e., Answer the Question, Topic of discussion, Summarise, Public Concerns). Leveraging the gpt-4o model, we formulate the answer by moulding the raw context into a specific format.

# 4 RESULTS AND DISCUSSION

As a case study, we implemented the VaxPulse QC by collecting the public discussion about the Shingrix® zoster vaccine for shingles. Australia's Therapeutic Goods Administrator approved Shingrix® zoster vaccine in 2021 but Shingrix® has been used internationally since 2017. The experiences and concerns of international consumers are helpful in knowing the public concerns and in formulating the educational material for Australians.

Online discussions on vaccination often cover various topics, including general discussions on shingles vaccines, the earlier Zostavax® vaccine, as well as concerns and adverse events experienced with co-administered vaccines. It requires segregating comments about the Shingrix® vaccine.

We segregated 35,103 vaccine-related comments from the 60,935 total comments posted on X (formerly Twitter), Reddit, YouTube and Facebook between January 2018 and October 2024.

## 4.1 Test Cases Generation

Generating hundreds of complex queries and their reference answers (ground truth) manually to evaluate the RAG pipeline is a time-consuming and laborious task. It is more difficult in our case because VaxPulse has four different categories of queries to extract more relevant information, and secondly, the reference answers have to be generated by processing thousands of comments. We adopt the Retrieval Augmented Generation Assessment (RAGAs) framework [17] for automated evaluation of our RAG pipeline.

RAGAs can also generate synthetic test cases to evaluate each component of an RAG pipeline. It uses a unique approach to generate questions with varying levels of difficulty, as well as different characteristics such as reasoning, conditioning, and multi-context, all derived from the provided dataset. This makes RAGAs an ideal tool for evaluating the performance of RAG pipelines. It not only generates the questions and their reference answers (also known as ground truths) but also extracts the context for the validation step. We generated test cases with a 50% ratio of simple questions and 25% for reasoning and multi-context questions, using default settings. However, we modified the test case generation prompt in RAGAs for the output options summarisation, public concerns, and topics of discussion. RAGAs generated a total of 276 test cases from the Shingrix dataset. This included 103 test cases for question/answer, 56 for extracting public concerns, 45 for summarising comments on vaccination, and 72 for generating discussion topics.

## 4.2 Results

The results of the VaxPulse QC evaluation for the context extraction component are presented in Table 2.

**Table 2: Context retrieval components evaluation results**

| Iteration | Context Precision | Context Recall |
|---|---|---|
| First Iteration (average) | 0.56 | 0.85 |
| Second Iteration (average) | 0.66 | 0.91 |
| Highest Scores | 1 | 1 |

Al Ghadban et al. [10] employed RAG to enhance healthcare information in low- and middle-income countries, assisting community health workers with maternal care through locally relevant information. They achieved scores of 0.16 in context relevancy and 0.26 in context recall.

VaxPulse's retrieval component improved from 0.56 to 0.66 for context precision and from 0.85 to 0.91 for context recall between iterations. The second iteration's context compression step, which extracts query-relevant parts of documents, contributed to this improvement. VaxPulse achieved perfect scores (1) for 13% of context precision and 90% of context recall test cases. Answers to the queries are evaluated based on their faithfulness and relevancy. Table 2 displays the evaluation results of the answer generation component.

Table 2: Answer Generation component evaluation

| Query Type | Faithfulness (Avg) | Answer Relevancy (Avg) |
|---|---|---|
| Answer the Question | 0.90 | 0.89 |
| Topics of Discussion | 0.90 | 0.91 |
| Summarise | 0.96 | 0.94 |
| Public Concerns | 0.88 | 0.82 |

The average faithfulness and relevancy scores of VaxPulse's answer generation component for the QA option are 0.90 and 0.89 respectively. The scores for extracting discussion topics are 0.90 and 0.91. VaxPulse's average score for the context summarisation is 0.96 and 0.94 which is due to the relevant context compression step in the second iteration. It achieved the highest score (1) for many test cases across different query types.

Social media is a mutually beneficial touchpoint for public health organisations and consumers to exchange information and promote relationships, especially for those with disabilities or chronic illnesses.

To effectively plan and communicate, health professionals must understand public sentiment and concerns. VaxPulse QC empowers stakeholders by providing tailored answers to specific queries. For instance, topic modelling might overlook localised concerns about the second dose of Shingrix® vaccine due to limited data. This tool can extract relevant comments to formulate informative responses. In conjunction with existing techniques in our framework (VaxPulse), this tool can assist the SAEFVIC team in identifying public concerns about an adverse event of special interest.

# 5 CONCLUSION

Social media is a rich source of public opinion, but analysing vaccine-related sentiment is challenging due to the sheer volume of data. Traditional methods like sentiment analysis and topic modelling often fail to capture nuanced opinions. VaxPulse QC, an efficient method, empowers public health administrators to understand vaccine concerns and hesitancies through customised queries. By leveraging the RAG pipeline, VaxPulse QC extracts key information from real-time data, mitigating the risk of LLM hallucinations.